# Extended point defects in crystalline materials: Ge and Si

N.E.B. Cowern[1,*], S. Simdyankin[1], C. Ahn[1,†], N.S. Bennett[1,‡], J.P. Goss[1], J.-M. Hartmann[2], A. Pakfar[3,§], S. Hamm[4], J. Valentin[5], E. Napolitani[6], D. De Salvador[6], E. Bruno[7], and S. Mirabella[7]

[1]School of Electrical and Electronic Engineering, Newcastle University, Newcastle upon Tyne, NE1 7RU, UK

[2]CEA, LETI, Minatec Campus, Grenoble, France

[3]ST Microelectronics, 850 rue Jean Monnet, 38920 Crolles, France

[4]Mattson Thermal Products GmbH, Daimlerstrasse 10, 89160, Dornstadt, Germany

[5]Probion Analysis, 37 rue de Fontenay, Bagneux, 92220, France

[6]CNR-IMM-MATIS and Dipartimento di Fisica, Università di Padova, Via Marzolo 8, 35131 Padova, Italy

[7] CNR-IMM-MATIS and Dipartimento di Fisica e Astronomia, Università di Catania, Via S. Sofia 64, 95123 Catania, Italy

**Abstract**

B diffusion measurements are used to probe the basic nature of self-interstitial 'point' defects in Ge. We find two distinct self-interstitial forms – a simple one with low entropy and a complex one with entropy ~30 k at the migration saddle point. The latter dominates diffusion at high temperature. We propose that its structure is similar to that of an amorphous pocket – we name it a *morph*. Computational modelling suggests that morphs exist in both self-interstitial and vacancy-like forms, and are crucial for diffusion and defect dynamics in Ge, Si and probably many other crystalline solids.

A vast array of crystalline material properties arises from the behavior of atomic-scale 'point' defects, yet these defects are poorly understood. Knowledge of simple point defects – single atoms added interstitially to, or missing from, an otherwise undisturbed lattice – is well established from quantum theoretical calculations and low-temperature experiments, but diffusion experiments hint that more complex entities may be involved at high temperatures relevant to industrial processing [1-5]. This Letter provides the first definitive evidence for these elusive complex defects and presents a specific physical model for their structure and diffusion.

Recent interest in Ge-based nano-electronics has led to basic studies on diffusion [5-9] and implantation defects [10,11] in crystalline Ge. Most dopants in Ge are found to diffuse by vacancy mechanisms, with activation energies below that of vacancy-mediated self-diffusion (≈ 3.1 eV), but boron diffusion is an exception with an activation energy of ≈ 4.65 eV [6,12]. Experiments [5,7-9] show that boron diffuses via the reaction B + I ⇔ BI, where 'B' represents substitutional boron, 'I' the self interstitial, and 'BI' a mobile dopant-interstitial complex. The energetics involved is illustrated in Figure 1.

The reduction in free energy on forming BI enables it to migrate a mean projected distance $\lambda$ before dissociating to B and I. The mean number of jumps before dissociation depends on the energy difference between migration and dissociation of BI and the diffusional entropies of I and BI. In general,

$$\lambda = \lambda_0 \exp(-E_\lambda / kT) \qquad (1)$$

where $E_\lambda = -(E_{self,X} + E_{barrier} - E_{AX})/2$ and $\lambda_0 = (4\rho a / f_{AX})^{1/2} \exp[(S_{AX} - S_{self,X})/2k]$, A is the impurity (here, boron), X the point defect driving AX diffusion (here, I), $a$ the capture radius for the forward reaction, $f_{AX}$ the diffusion correlation factor (~1), $E_{AX}$, $S_{AX}$, $E_{self,X}$, $S_{self,X}$ the activation energies and entropies of impurity diffusion and self-diffusion via the species AX and

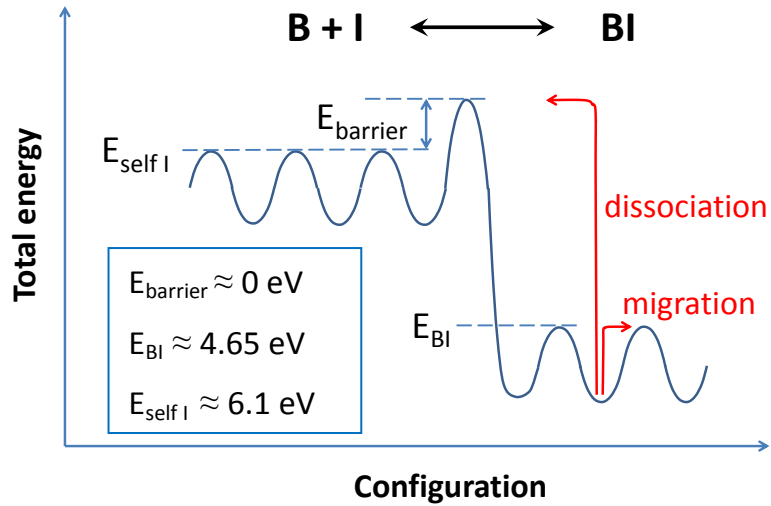

**Fig. 1.** Schematic of total energy versus configuration for the reaction mediating B diffusion in Ge. Also shown are energies inferred from previous experiments. $E_{BI}$ and $E_{self,I}$ are the respective energies of BI and I at their migration saddle points, relative to that of substitutional B.

X, respectively, and $E_{barrier}$ the energy barrier to the forward reaction (Fig. 2). Similarly to the case of Si, where $E_\lambda = -0.5$ eV [13] and $E_{barrier} < 0.05$ eV [14], recent experiments in Ge in the temperature range $T > 0.65\, T_m$ yield $E_\lambda$ in the range $-0.8$ eV [7] to $-0.6$ eV [8] with $E_{barrier} \approx 0$ eV [7-9]. Using equation (1) above this implies $E_{self,I} - E_{BI} \approx 1.2$ to $1.6$ eV, and since $E_{BI} \approx 4.65$ eV [6], we find $E_{self,I} \approx 5.85$ to $6.25$ eV. This is nearly 2 eV higher than predicted from first principles [15], and more than 1 eV higher than $E_{self,I}$ in Si [16].

At lower temperatures, data on B diffusion in Ge show a quite different behavior of the migration length [7-9]; as Figure 2 shows, $E_\lambda$ changes from about $-0.75$ eV at $T > 550°C$ to about $+0.06$ eV at lower $T$. It has previously been suggested that the high-$T$ results reflect the true activation energy while the low-$T$ results are an artifact of BI trapping at C or O atoms in the MBE-grown Ge used in Ref. 7. Here we propose an alternative view; the low-$T$ results arise from dissociation into a different self-interstitial species with lower activation energy and entropy.

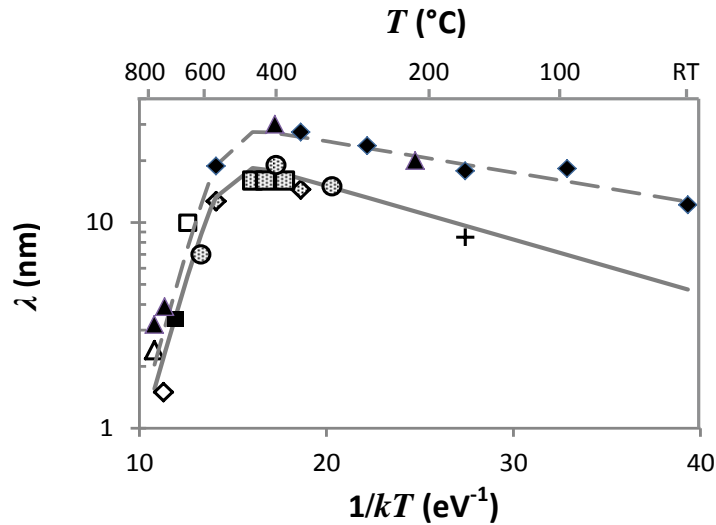

**Fig. 2.** Migration length of BI in Ge versus $1/kT$. Open symbols: thermal diffusion data from Refs. 7 (diamond) and 9 (triangle), and oxide precipitate-enhanced diffusion data from Ref. 17 (square). Shaded symbols: post-implant diffusion data from this work (circles), Ref. 7 (diamonds), and Ref. 8 (squares). Solid symbols: H-irradiation enhanced diffusion (RED) data from Ref. 7 (diamonds), Ref. 9 (triangles), and Ref. 17 (square). Cross symbol: O-RED data from Ref. 18 (O gives less ionization than H per atomic displacement). The curves are fits of equation 1 generalized to account for competing dissociation channels to two self-interstitial forms, $I$ and $\mathcal{I}$. The best fit under non-irradiation conditions (solid curve) is obtained with $E_\lambda = (-0.725 \pm 0.10)$ eV, $\lambda_o = 0.62$ pm for $\mathcal{I}$, and $E_\lambda = (0.06 \pm 0.02)$ eV, $\lambda_o = 50$ nm for $I$. Under RED conditions (dashed curve) the fitted values of $E_\lambda$ shift 0.025 eV in the negative direction. This could be accounted for by a reduction of 0.05 eV in the migration energy of BI under H irradiation.

To test this idea we have repeated the experiments, now using CVD-grown epitaxial Ge in which trap concentrations are definitively too low to affect B diffusion [12]. Extracted values of $\lambda$ are shown in Fig. 2, together with previous results [7-9,17,18]. Our data points are perfectly consistent with the earlier results, despite the absence of traps – thus strongly supporting our proposal of two self-interstitial species. Assuming $E_{BI} = 4.65$ eV over the full temperature range of Fig. 2, we have fitted the results in Figure 2 with a formula based on two self-interstitial forms [12]. For the low-$T$ self-interstitial, which we label $I$, we find $E_{\text{self},I} \approx 4.55$ eV and $S \approx 4$ k, whereas for the high-$T$ self-interstitial, labeled $\eta$, $E_{\text{self},\eta} \approx 6.1$ eV and $S \approx 30$ k. The low-$T$ value of $E_{\text{self},I}$ agrees with first-principles calculations for a localized self-interstitial in Ge [4] and the corresponding entropy value confirms $I$ is indeed a simple point defect. In contrast, the high-$T$ value is unexplained by theory and its entropy has an extreme, record-breaking value. One way to explain this would be to invoke pre-melting effects, i.e. melting fluctuations that occur close to the transition to the liquid phase. However, this explanation seems to be ruled out by the near-constant activation energy over the observed temperature range, and the fact that this range is far below the melting point. We propose instead that $\eta$ has a complex, thermodynamically stable structure incorporating a number of atoms from the lattice.

The sharpness of the transition between the two diffusion regimes can be explained if there is a reaction barrier between the two defect forms. Fig. 3 shows a model for the energy and entropy of self-interstitials that extend over different volumes of the lattice ($N$ atoms occupying a volume normally occupied by $N - 1$ lattice atoms). At low temperature the simple form is dominant, while at high temperature the complex form dominates.

It is obviously of great interest to know what physical form the complex defect takes, and a simple model of a small disordered region leads to interesting semi-quantitative predictions. A rough upper limit on the number of atoms in the defect is $N < S_{\text{self},I}/s_f$, where $s_f$ is its formation entropy per atom and the inequality applies because $S_{\text{self},I}$ includes both formation and migration entropy. Applying this to Seeger's 'liquid drop' model of an extended point defect [1], $s_f$ would be the entropy of fusion, 3.6 k/atom, resulting in a value of $N < 8$. This is too small to behave like a bulk liquid as confined liquids become solid-like [19], with much lower entropy and internal energy per atom.

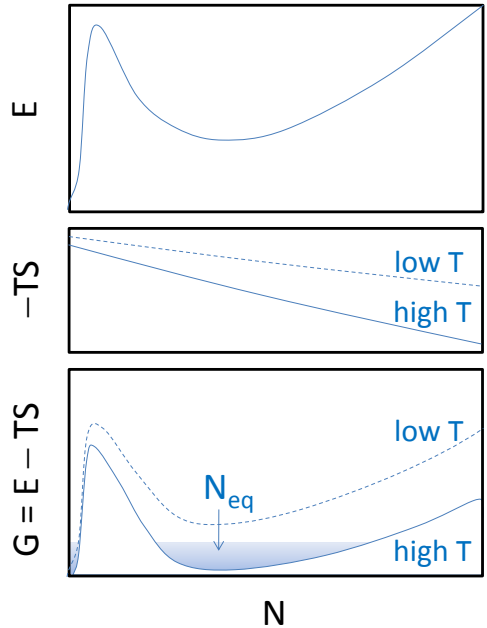

**Fig. 3:** Schematic diagram of enthalpy, $E$, entropic energy term, $-TS$, and resultant Gibbs free energy, $G$, for the self-interstitial as function of size, illustrating how a simple structure may dominate at low $T$ and a complex structure dominate at high $T$. $N$ represents the number of lattice atoms incorporated into the defect; zero or one in the case of a compact self-interstitial (e.g. a simple or split interstitial), but much larger in the case of an amorphous pocket, though this still only contains one excess atom.

Thus it is interesting to consider a larger structure, with a formation energy per atom rather close to that of the crystalline solid, yet with high entropy. We postulate an extended region, or *morph*, having regular coordination with the surrounding lattice but containing the basic building blocks found in amorphous material (for example, in Ge and Si, four, five, six and/or seven-membered rings). In the case of an interstitial-like defect (i-morph) the structure would contain one extra atom and in a vacancy-like defect (v-morph) there would be a deficit of one atom.

A very rough estimate of the formation energy of morphs, independent of specific structures, can be obtained using a semi-empirical, macroscopic approach. We assume a spherical inclusion and write

$$E_f = H_c N + E_\varepsilon(\mu) + E_{bd}; \qquad \mu = (1+\alpha)(N+n)/N - 1$$

where $H_c$ is the heat of crystallization of the amorphous phase, $E_\varepsilon$ is the misfit strain energy, and $E_{bd}$ is an additional 'bond distortion energy' [12]. Literature values for $H_c$ are significantly scattered [20,21] and not always mutually consistent with published entropy values – here we use $H_c \approx 0.1$ eV/atom for Ge and 0.12 eV/atom for Si. The misfit $\mu$ contains two factors, one related to the volume mismatch $\alpha$ between crystalline and amorphous phases ($\alpha \approx 1.5\%$ for Si and Ge), the other to the excess number of atoms in the defect ($n = 1$ for ℐ, $n = -1$ for 𝒱, the v-morph). Strain energy is roughly estimated from the Birch-Murnaghan high-pressure equation of state [22], assuming the defect is surrounded by a rigid matrix [12]. Finally, we choose $E_{bd} = 1$ eV to match our measured activation energy for ℐ in Ge. This procedure yields rough estimates of $E_f$ for ℐ and 𝒱 in Ge and Si as shown in Fig. 4, panels (a), (e). Energy is minimized when strain energy (which decreases with $N$ at small $N$ values) equals constitutive energy (which increases monotonically with $N$).

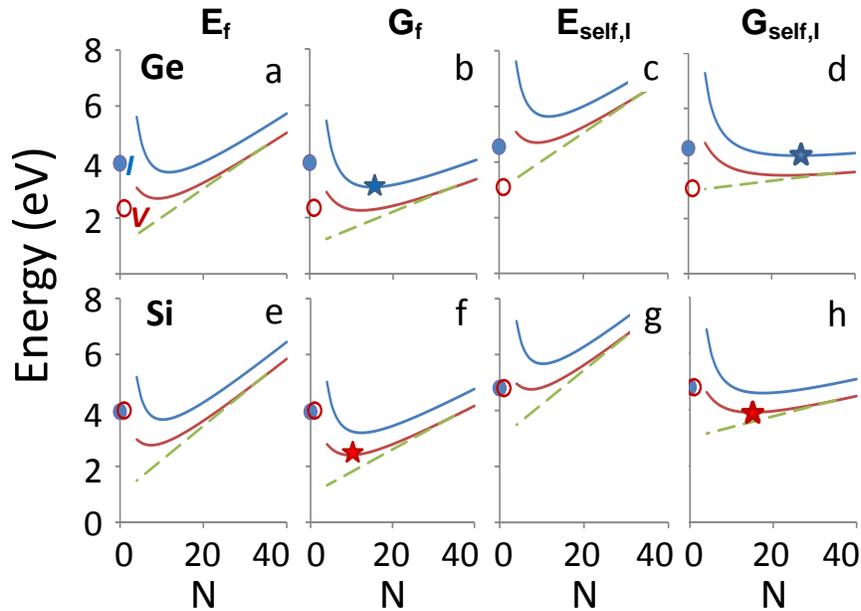

**Fig. 4.** Energy and Gibbs free energy (at 700°C) of defect formation (left-hand panels; a, b, e, f) and self diffusion (right-hand panels; c, d, g, h) for Ge (top) and Si (bottom), as functions of $N$. Values for simple point defects are shown at left of each panel ($I$ – closed symbol; $V$ – open symbol). Estimated values for morphs are shown by solid curves (ℐ – upper curve, 𝒱 – lower curve). Stars indicate the free-energy minima for morphs in their stable state (**b**, **f**), and at the diffusion saddle point (**d**, **h**). Dashed lines show values when strain energy is omitted.

Since entropy increases with $N$, at finite $T$ the minimum in Gibbs free energy occurs at larger $N$. For ϑ, an entropy of 1 $k$/atom in Si or Ge (assumed equally distributed between formation and migration entropy) and a value of $N \approx 30$ in Ge gives good agreement with our experiment-based diffusion entropy estimate of 26.6 $k$. For the sake of precision it is worth noting that, during morph migration the center of mass of the defect moves by only a fraction of the normal self-interstitial jump length per rebonding event [12]. This leads to a small correction of about 2.5 $k$ in the entropy inferred from our experiment, i.e. we obtain $S \approx 30\ k$. However, as our model is inherently approximate and our estimated diffusion entropy per atom is drawn from scattered literature data [23] our results should not be seen as exact predictions but rather as an indication of trends.

We now take a closer look at the trends shown in Fig. 4, and draw comparisons with published experimental data where available. As shown in the left half of Fig. 4, the predicted formation energies of morphs are comparable to those of simple point defects and their Gibbs free energies are lower. This suggests the predominant vacancy and self-interstitial species in Si might also be morphs. Low $G_f$ values for morphs could account for numerous unexplained experimental observations. For example, the huge scatter of literature data for point-defect concentrations and diffusivities in Si could arise from different coupling between populations of low-concentration, fast-diffusing simple point defects and high concentration, slow-diffusing morphs in different experiments. This interplay could be crucial for understanding defect formation during crystal growth and electronic device fabrication, and explain discrepancies between point-defect parameters needed to model processes at different length, time and temperature scales.

The relative contributions of simple point defects and morphs to diffusion also depend on their respective migration energies. Morph migration relies on peripheral rebonding, the process involved in solid-phase epitaxy (SPE). Hence, for a rough estimate of morph migration energies we use the SPE activation energies for Ge [24] and Si [25]; 2.1 eV and 2.85 eV, respectively. As shown in the right half of Fig. 4, for ϑ this leads to $E_{self}$ values of around 6 eV in Si and Ge, but a significantly lower $G_{self}$ owing to the large entropies involved.

In the case of 𝒱 we find a self-diffusion activation energy of about 5 eV in Si and Ge. This indicates the compact vacancy, with activation energy 3.1 eV, dominates self-diffusion in Ge, a result consistent with experimental data showing a single activation energy over several hundred degrees [26]. However, in Si, our model suggests 𝒱 may contribute significantly. Our estimated 5 eV activation energy is close to the experimental value of 4.86 eV for vacancy-mediated self-diffusion in Si at high $T$ [4]. Moreover, our high migration-energy value of 2.85 eV for the morph in Si agrees quite well with the vacancy migration energy observed experimentally at high $T$ [3] ($\approx$ 1.65 eV, contrasting with $\approx$ 0.5 eV at low $T$). This consistent picture of a morph-type vacancy at high $T$ and a simple vacancy at low $T$ resolves previous controversy on vacancy-mediated self-diffusion [27]. This point is discussed in further detail in Ref [12].

It is also instructive to compare our modeled entropy values with published experimental data. Our values of $\approx$ 9 $k$ for 𝒱 in Si, 16 $k$ for ℐ in Si, and 30 $k$ for ℐ in Ge are well matched by experimental high-$T$ diffusion entropy values of 10 $k$ for 𝒱 in Si [4], 12 $k$ for ℐ in Si [4], and about 30 $k$ for ℐ in Ge (our experiments), respectively (star symbols in Fig. 4, panels (f) and (d), again attributing 1 $k$/atom).

Finally, our model may resolve several further unexplained features of diffusion and defect dynamics in Si and Ge which at first sight would seem unrelated. First, it predicts broadly similar free energies of formation, migration, and thus self-diffusion for ℐ and 𝒱 in the same material, because the free energy of a morph is related to the number of atoms involved. For example, a morph with 29 atoms on a 30-atom crystalline 'footprint' is a 𝒱, while one with 31 atoms is an ℐ; these will have similar configurational free energies of formation and migration. This neatly explains, for the first time, the notable 'coincidence' of interstitial and vacancy-mediated self-diffusion coefficients in Si at high $T$ [4]. Second, the predominance of a complex, high-entropy self-interstitial in Ge may explain the anomalously low recombination rate for self-interstitials at the Ge surface [5]. If ℐ recombines at specific localized sites, as is thought to occur with simple point defects, recombination will be inhibited by a free-energy barrier as the defect shrinks and annihilates. Third, the low formation energies of interstitial and vacancy-type morphs in Ge may explain the ease with which Ge amorphizes during ion bombardment with energy density above about 0.1 eV/atom [28]. In this picture, rather than forming predominantly simple point defects that can migrate and recombine, the bombardment produces ℐ- and 𝒱-morphs that are immobile at room temperature and so accumulate and ultimately overlap.

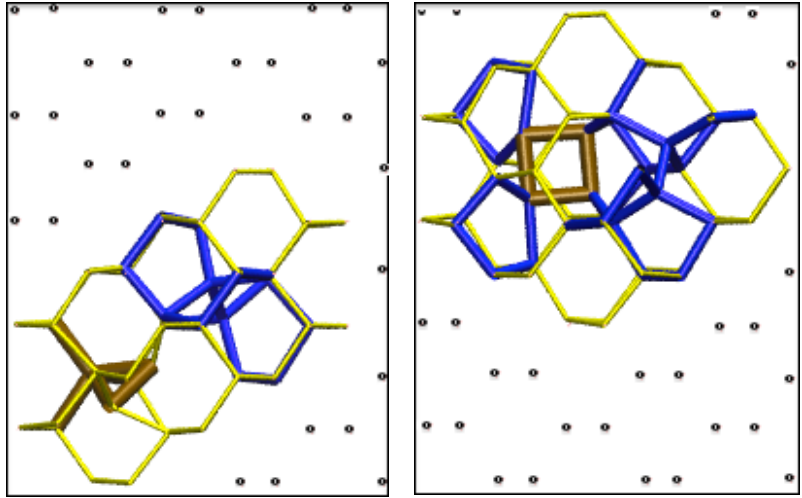

**Fig. 5.** Two MD simulation frames, separated by 20 ps, showing migration of an i-morph through part of a 10,000 atom simulation volume. Seven-membered rings are shown with thin (yellow) 'bonds', 5-membered with thicker (blue) bonds, and fewer-membered with the thickest bonds. Dots mark atomic positions on the surrounding diamond lattice.

It is clearly important to test the predictions of our semi-empirical model against atomistic calculations. We have therefore conducted initial molecular dynamics (MD) calculations, using a potential that gives an energy gap between the crystalline and relaxed amorphous phases of about 0.1 eV [29], close to experimentally observed values for Si and Ge. This choice is crucial to success as substantially higher energy gaps ($\approx$ 3 times higher with the frequently used Stillinger-Weber potential [30]) incorrectly penalize morph formation. We find characteristic morph structures that are thermally stable and mobile at high temperature, and migrate by shape-shifting through numerous configurations of similar energy. By way of illustration, two snapshots of the same 'self-interstitial' defect, taken 20 ps apart during diffusion at $T \approx 0.95\ T_m$, are shown in Fig. 5. At lower temperatures similar morph structures are seen and occasional transitions between the compact and morph forms occur. Broadly similar results are found for vacancies, with a slightly smaller defect size and fewer structural permutations – again consistent with our predicted trend. A full discussion of MD calculations for native defects in the diamond lattice at high temperature will be presented elsewhere.

In conclusion, at high temperature the self-interstitial in Ge is a complex, mutable yet robust structure of dimensions ~1 nm, with a structure similar to an amorphous pocket. Analogous morph structures are expected to exist for both the self-interstitial and vacancy in Si. More generally, there is the exciting possibility that morphs occur throughout the wide range of crystalline materials that have a small amorphous-crystalline energy gap, including important geophysical materials like ice and advanced technological materials such as high-κ dielectrics. Since point-defect properties are fundamental to materials behavior in applications from industrial processing to glacier dynamics we believe this novel class of point defects merits extensive further study.

**Acknowledgments:**

We are grateful to Paul Coleman and Charlene Edwardson for carrying out positron annihilation measurements, to David Drabold and Yuting Li for sending us some of their amorphous Si models, to Normand Mousseau and Ali Kerrache for sharing their Fortran code, which we modified for the ring analysis used in Fig. 5, and to Robert Falster and Vladimir Voronkov for illuminating discussions. The research leading to these results received funding from the European Union Seventh Framework Programme (FP7/2007-2013) under grant agreement no. 258547 (ATEMOX).

\*   Contact author e-mail: nick.cowern@ncl.ac.uk.

†   Now at Samsung Semiconductor Inc., 75 W Plumeria Dr., San Jose, CA 95134.

‡   Now at Nanomaterials Processing Laboratory, School of Electronic Engineering, Dublin City University, Glasnevin, Dublin 9, Ireland.

§   Now at GLOBALFOUNDRIES, Wilschdorfer Landstrasse 101, 01109 Dresden, Germany

**References**

1. A. Seeger, Phys. Stat. Sol. B **248**, 2772 (2011).

2. A. Seeger. Radiation Effects **9**, 15 (1971).

3. H. Bracht *et al*., Phys. Rev. Lett. **91**, 245502 (2003).

4. A. Ural, P.B. Griffin, and J.D. Plummer, Phys. Rev. Lett. **83**, 3454 (1999).

5. H. Bracht *et al*., Phys. Rev. Lett. **103**, 255501 (2009).

6. S. Uppal *et al*., J. Appl. Phys. **96**, 1376 (2004).

7. E. Bruno *et al*., Phys. Rev. B **80**, 033204 (2009).

8. E. Napolitani *et al*., Appl. Phys. Lett. **96**, 201906 (2010).

9. G.G. Scapellato *et al*., Nucl. Instrum. Meth. Phys. Res. B **282**, 8 (2012)

10. G. Bisognin, S. Vangelista, and E. Bruno, Mat. Sci. Eng. B **154-155**, 64 (2008).

11. A. Claverie, S. Koffel, N. Cherkashin, G. Banassayag, P. Scheiblin, Thin Solid Films **518**, 2307 (2010).

12. See Supplemental Material (below)

13. N.E.B. Cowern, K.T.F. Janssen, G.F.A. van de Walle, and D.J. Gravesteijn, Phys. Rev. Lett. **65**, 2434 (1990).

14. N.E.B. Cowern *et al*., Mat. Sci. Semicond. Process. **2**, 369 (1999).

15. C. Janke, R. Jones, S. Öberg, and P.R. Briddon, J. Mater. Sci.: Mater. Electron. **18**, 775 (2007).

16. H. Bracht, E.E. Haller, R. Clark-Phelps, Phys. Rev. Lett. **81**, 393 (1998).

17. G.G. Scapellato, *B and Sb in germanium for micro and optoelectronics*, PhD thesis, University of Catania (Catania, Italy, Dec. 2011).

18. E. Bruno *et al*., Thin Solid Films **518**, 2386 (2010).

19. J. Gao, W.D. Luedtke, and U. Landman, Phys. Rev. Lett. **79**, 705 (1997).

20. A. Gabriel, *Atomistic simulation of solid-phase epitaxial regrowth of amorphous germanium*, PhD Thesis, Technical University of Dresden (2008).

21. F. Kail *et al*., Physica Status Solidi *(RRL)* **5**, 361 (2011).

22. F. Birch, Phys. Rev. **71**, 809 (1947).

23. R.L.C. Vink and G.T. Barkema, Phys. Rev. Lett. **89**, 076405 (2002).

24. D.Y.C. Lie, J. Electron. Mater. **27**, 377 (1998).

25. M. Bauer *et al*., Thin Solid Films **364**, 228 (2000).

26. G. Vogel, G. Hettich, and H. Mehrer, J. Phys. C **16**, 6197 (1983).

27. G.D. Watkins, J. Appl. Phys. **103**, 106106 (2008).

28. C. Claeys and E. Simoen, in *Extended defects in germanium: Fundamental and technological aspects*, Ch. 5, 241–292 (Springer Series in Materials Science, Vol. 118, 2009).

29. J. Tersoff, Phys. Rev. B **37**, 6991 (1988).

30. M.J. Caturla, T. Diaz de la Rubia, and G.H. Gilmer, Nucl. Instrum. Meth. Phys. Res. **106**, 1 (1995).

**Supplemental Material:**

*Experiments and analysis*

Our experiments used Ge self-implantation into CVD grown single-crystal Ge – material which is expected to contain much lower C and O concentrations than those in previous experimental studies – too low to cause saturation at the observed migration length. Ge containing a B-doped marker with a peak concentration of about $10^{19}$ cm$^{-3}$ at a depth of about 400 nm was grown epitaxially on a Si wafer using a Ge/graded SiGe virtual substrate. This structure was implanted with 60 keV Ge ions to a dose of $1 \times 10^{15}$ cm$^{-2}$, producing a near-surface amorphous layer and an underlying interstitial-rich layer of damaged crystalline Ge. On annealing, the amorphous layer re-grew by solid phase epitaxy, leaving a band of interstitial-type 'end-of-range' defects embedded in the crystalline Ge epilayer. Samples were annealed at temperatures in the range 300 to 600°C. Throughout this range transient enhanced diffusion of B was observed in the buried marker layer, originating from the emission and indiffusion of self-interstitials from the end-of-range band. B diffusion was measured by secondary-ion mass spectrometry (SIMS) of annealed and unannealed samples, and nonlinear least-squares data fitting was used to extract migration lengths and migration frequencies as in Refs. 8 and 13. Values of $\lambda$ extracted from this analysis are shown in Fig. 2 of the paper together with previous results. The corresponding numerical values and source publications are listed in Table S1 below.

Our data obtained using CVD Ge are close to those previously obtained using MBE Ge. This suggests that the model of trapping of BI by impurities, previously suggested in Ref. 8, may not be the correct explanation for the dramatic change in slope of $\lambda$ versus $1/kT$ shown in Fig. 2 of our paper. However, we first need to rule out unexpectedly high C or O concentrations in our CVD material, and to do this we have performed SIMS measurements for these impurities. Concentrations of both impurities are below the SIMS sensitivity of $\sim 2 \times 10^{16}$ cm$^{-3}$ for C and $3 \times 10^{16}$ cm$^{-3}$ for O. We have also investigated the (unlikely) possibility of a high concentration of open volume defects in epitaxial Ge, which might act as alternative traps for BI. Positron annihilation spectroscopy measurements showed no evidence for vacancy-type defects in the region of the B-doped marker at the detection level of a few times $10^{16}$ cm$^{-3}$.

Diffusion data also give a positive indication of low trap content in our material, in that self interstitials are able to diffuse with little trapping from the EOR defect band to the location of the marker layer, about 340 nm deeper into the Ge. This suggests an effective trap concentration below $10^{14}$/cm$^2$.

Finally, in relation to the published results using MBE Si, data in Fig. 3.22 of Ref. 17 appear to indicate multiple as well as single migration events, even when the extracted value of $\lambda$ is close to the apparent limiting trapping length. This again supports the view that the measured migration length is not caused by trapping, but by dissociation, BI → B + I, leading to further migration events via B + I → BI.

### $\lambda$ in the presence of two dissociation channels

The migration length of a diffusing species AX formed by the reaction A + X → AX between a substitutional impurity A and an intrinsic point defect X, can be expressed as $\lambda = \sqrt{D_{AX}/r}$, where $r$ is the dissociation rate to form the separated species A and X[13]. When the intrinsic defect has two possible forms, $X_1$ and $X_2$, each of which can react with A to form AX, the migration length generalizes to $\lambda = \sqrt{D_{AX}/(r_1 + r_2)}$, where $r_i$ is the dissociation rate leading to species $X_i$. Defining $\lambda_1 = \sqrt{D_{AX}/r_1}$ and similarly for $\lambda_2$, we can write $1/\lambda^2 = 1/\lambda_1^2 + 1/\lambda_2^2$. The parameters $\lambda_1$ and $\lambda_2$ are defined as in equation (1) of the paper. In the case discussed in the paper, $X_1$ represents the compact self-interstitial $I$ and $X_2$ represents the extended self-interstitial $\mathcal{I}$.

*B diffusivity in Ge*

Data on the temperature dependent diffusivity of B in Ge have been reported by Uppal et al.[6] in the temperature range from 800 - 900°C. Here we extend the measurement range to lower temperatures using long time anneals (up to 96 hours) of MBE-grown B marker layers at temperatures down to 720°C.

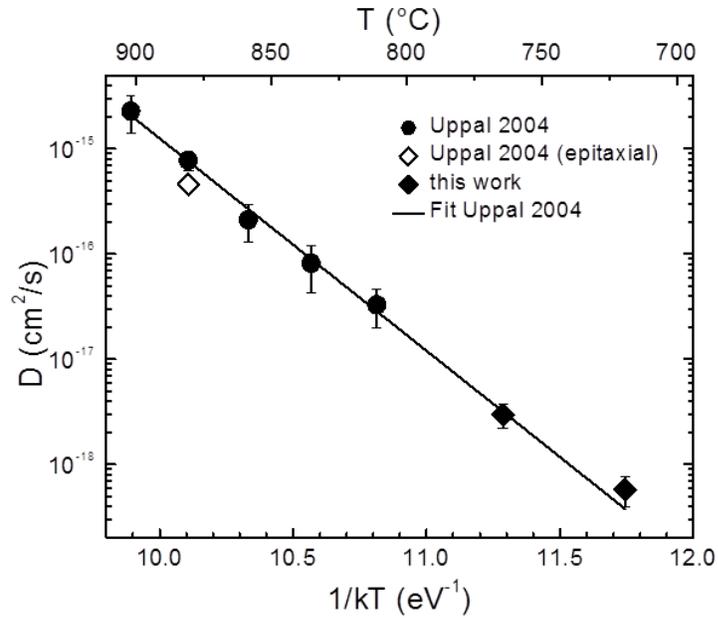

Fig. S1: Data on the B diffusion coefficient in Ge as a function of temperature from Uppal (ref. 6) and this work.

The data are consistent, within experimental error, with Uppal's results which gave an activation energy of 4.65 eV and we therefore use Uppal's result with increased confidence for the analysis in this paper.

*Bond distortion effect on morph formation energy*

In our simple semi-empirical model for the formation energy of a morph, the leading terms are constitutive energy and strain energy. The first reflects the energy difference between the amorphous and crystalline phases, including the inherent bond distortion energies, both from bond-length and bond-angle terms, of the relaxed amorphous phase. The second is a continuum estimate of 'long range' strain arising from the volume mismatch between the defect and its surroundings. A third term may be required to take account of additional bond distortions at the microscopic level, arising from the attachment of bonds across the interface between the morph and the surrounding crystal. These distortions affect bonds inside the defect up to about a distance $L$ from the interface, where $L$ is the correlation length of the amorphous system. Typically, $L$ is comparable to the size of the morph, so most bonds in the defect are involved. This leads to an additional term in the formation energy.

In the case of $\eta$ in Ge, the constitutive and strain energy terms appear to dominate, leaving only a 'missing' ~1 eV to account for our observed experimental result. This suggests that the additional local bond distortion energies are small. Noting that the additional energy per bond will decrease as the defect size increases, while the total number of bonds increases, we assume that these trends roughly cancel, producing a constant contribution to the formation energy (the term $E_{bd}$ which we set to 1 eV). This crude assumption can be improved upon in future with the help of atomistic calculations. Meanwhile we suggest our estimate is reasonable down to defect sizes where the morph structure fails due to topological constraints.

*Effective jump length for morph diffusion*

In our simplified model of morph diffusion an (approximately spherical) morph migrates by successive rebonding events at the interface between it and the surrounding crystal. Figure S2 shows schematically the result of one diffusion 'jump', consisting of a rebonding event where the periphery of the morph moves out to incorporate one additional atom from the lattice. Since one atom has been added, the centre of gravity of the defect has shifted by a distance $l_a \approx r_a / N$, where $r_a$ is the defect radius. This is the effective jump distance for the morph. The volume of a spherical morph on a footprint equivalent to $N$ atoms of the crystal lattice is $\frac{4}{3}\pi r_a^3 = N/\rho_A$ where $\rho_A$ is the number of atoms per unit volume of crystal. Hence $r_a = (3N/4\pi\rho_A)^{1/3}$ and the jump distance $l_a \approx (\frac{4}{3}\pi\rho_A N^2)^{-1/3}$. In the case of the diamond lattice of Si or Ge, we have $\rho_A = (2/a)^3$, where $a$ is the lattice constant, and consequently $l_a = (a/2)/(\frac{4}{3}\pi N^2)^{1/3}$. This is substantially smaller than the usual jump length for a self-interstitial, and thus slightly modifies the entropy to be inferred from diffusion data, by ~2.5 $k$. This implies that the inferred entropy is model-dependent, but in practice the effect is small and does not affect our conclusions.

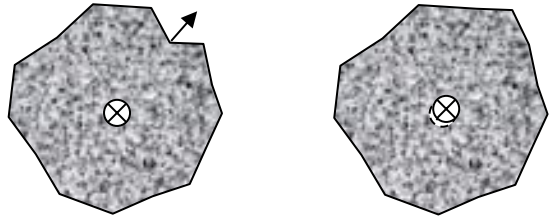

**Fig. S2.** Schematic of morph before (left) and after (right) a rebonding event that moves the defect boundary. The centre of gravity is indicated by the crossed circle. After rebonding the centre of gravity of the morph has moved a small distance (dashed circle shows its initial position). This is the migration jump length for a morph – significantly lower than the usual jump length for a simple point defect. A calculation of the jump length is given in the above text.

*Controversy over the role of complex point defects in vacancy-mediated self diffusion in Si*

In 1999 Ural *et al*.[4] conducted experiments on Sb, P, and Si isotope diffusion in CVD-grown Si under injection of self interstitials or vacancies in the temperature range 800 – 1100°C. Their results showed that interstitials and vacancies, or some form thereof, contribute roughly equally to self-diffusion in this temperature range, in this material. In 2007 Shimizu and coworkers [Shimizu, Y., Uematsu, M., Itoh, K.M., Experimental evidence of the vacancy-mediated silicon self-diffusion in single-crystalline silicon, *Phys. Rev. Lett.* **98**, 095901 (2007)] conducted experiments on self diffusion in MBE-grown Si at temperatures from 735 to 875°C. Their diffusion measurements were performed only under equilibrium point-defect conditions (no vacancy or interstitial injection), and so they were unable directly to distinguish between interstitial and vacancy components. Nevertheless, they assumed that their results represented vacancy-mediated diffusion as the slope of their data on an Arrhenius plot gave an activation energy of 3.6 eV compared to the 4.95 eV found by Bracht at higher temperatures. Watkins later pointed out[27] that the results of Shimizu *et al.* agreed with an extrapolation from independent data obtained at cryogenic temperatures, where the vacancy is known to be a simple point defect.

Watkins further argued that, since a simple vacancy could explain the results of Shimizu, there was no need to invoke a more complex vacancy in silicon at elevated temperatures. This led him to conclude that Bracht's estimate of a large vacancy migration energy at high temperature, associated with a complex vacancy[3], was probably wrong. However, this argument overlooked the results of Ural et al.[4], which had clearly shown that a vacancy of some kind, with a higher activation energy and entropy, is involved in self diffusion at high temperatures.

Our present paper resolves this apparent contradiction in the literature by pointing out that at 'low' temperatures in Si (up to the 800°C range, thus including the range that Watkins referred to in Ref. 27 as 'elevated temperatures'), the dominant vacancy is $V$, while at higher temperatures the dominant vacancy is $\mathcal{V}$. Incidentally, this temperature is ~0.65 $T_m$ – the same fraction of the melting point at which interstitial-mediated self-diffusion in Ge becomes dominated by $\mathcal{I}$.

Existing data on interstitial-mediated self-diffusion in Si suggest a similar temperature-dependent transition, although the effect is closer to the limits of experimental uncertainty. Data obtained from the analysis of B diffusion during self-interstitial cluster ripening show an activation energy of 4.75 eV for interstitial-mediated self-diffusion in the temperature range 600 – 800°C [Cowern, N.E.B. *et al.*, Cluster ripening and transient enhanced diffusion in silicon. Mat. Sci. Semicond. Proc. **2**, 369-276 (1999)]. They cannot be satisfactorily fitted using the 4.95 eV determined by Bracht at temperatures between 800°C and the melting point. Interestingly, again, the break point between the two sections of data occurs at around 0.65 $T_m$, suggesting that the transition from simple to complex point defects mediating self-diffusion occurs at about the same temperature for vacancies and interstitials. This analysis indicates that experimental results at high temperature, obtained in Refs. 3, 5 and related publications, reflect the behaviour of morphs and not that of simple point defects.

*Entropy of small morphs*

For defect sizes below a threshold value $N_{crit}$, it becomes impossible to 'fit' typical subunits of an amorphous network into the surrounding crystal without forming energetically expensive structures such as dangling bonds. Such structures may act as an energetic barrier between the morph and simple forms of a point defect. At marginally larger sizes, only a few structures will be available with energy per atom comparable to amorphous silicon. Entropy in this size range may be smaller than naïve expectation, based on the number of atoms in the defect, would suggest.

Based on this argument a current best estimate of $N_{crit}$ for Si or Ge can be derived, somewhat speculatively, from experiment. First, our result for ? in Ge shows a very high entropy defect, consistent with a morph with size $N \sim 30$, having many structural or vibrational degrees of freedom. Second, published diffusion data[3,5], together with our model, suggest that interstitials and vacancies at high $T$ in Si are morphs of size $N \sim 10$, but the result of Bracht *et al.*[3] suggests that the associated formation entropy might be small, $(0 \pm 6)$ k. Taken at face value, this suggests that the vacancy in Si is only slightly larger than $N_{crit}$, as illustrated schematically in Fig. S3. Since we have inferred $N \sim 10^1$ for the vacancy morph in Si, it is reasonable to speculate that $N_{crit} \sim 10^1$. This conclusion is supported by molecular dynamics simulations of vacancy morphs in Si (to be published) which confirm that they are smaller (N < 20) and have fewer possible configurations than self-interstitial morphs.

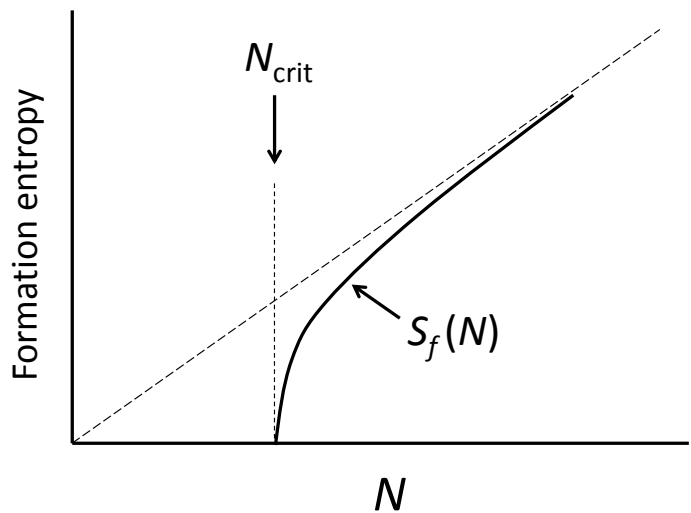

**Fig. S3.** Schematic of the size dependent formation entropy of an morph. Topological constraints for an amorphous-like structure lead to a threshold size, $N_{crit}$, and a reduction in the entropy of defects just above this size. This may explain why the vacancy in Si exhibits the surprising combination of large migration energy, typical of a morph-like structure, with low formation entropy, more usually associated with a compact point defect.

**Table S1.** Numerical values of λ (nm) used in Fig. 2 of the paper. H-RED - hydrogen irradiation enhanced diffusion, O-RED - oxygen irradiation enhanced diffusion, TD - thermal diffusion, TED - post-implant transient enhanced diffusion, OED – oxygen precipitate-enhanced diffusion. TD, TED, OED are all non-irradiation conditions.

| 1/kT (eV$^{-1}$) | H-RED Ref. 7 | TD Ref 7 | TED Ref. 7 | TED Ref. 8 | TED This work | H-RED Ref. 17 fig. 3.24 | OED Ref. 17 fig. 3.22 | TD Ref. 9 | H-RED Ref. 9 | O-RED Ref. 18 |
|---|---|---|---|---|---|---|---|---|---|---|
| 10.80 | | | | | | | | 2.4 | 3.2 | |
| 11.29 | | 1.5 | | | | | | | | |
| 11.34 | | | | | | | | | 3.9 | |
| 11.93 | | | | | | 3.4 | | | | |
| 12.57 | | | | | | | 10.0 | | | |
| 13.29 | | | | | 7.0 | | | | | |
| 14.10 | 18.8 | | 12.7 | | | | | | | |
| 16.05 | | | | 16. | | | | | | |
| 16.74 | | | | 16. | | | | | | |
| 17.24 | | | | | | | | | 30. (re-analysis) | |
| 17.30 | | | | | 19. | | | | | |
| 17.77 | | | | 16. | | | | | | |
| 18.62 | 27.5 | | 14.5 | | | | | | | |
| 20.3 | | | | | 15. | | | | | |
| 22.18 | 23.6 | | | | | | | | | |
| 24.75 | | | | | | | | | 20. (re-analysis) | |
| 27.42 | 17.8 | | | | | | | | | 8.5 |
| 32.86 | 18.3 | | | | | | | | | |
| 39.32 | 12.2 | | | | | | | | | |